\documentclass[%
 reprint,
 showpacs,preprintnumbers,
 amsmath,amssymb,
 aps,
 prb,
]{revtex4-2}
\usepackage{dcolumn}
\usepackage{subfigure}
\usepackage{graphicx,color}
\usepackage{mathrsfs}

\begin{document}

\title{Optical conductivity of the topologically-nontrivial MXenes, Mo$_2$HfC$_2$O$_2$ and W$_2$HfC$_2$O$_2$: first-principles calculation and effective model analysis}

\author{Tetsuro Habe}
\affiliation{Department of Mechanical and Electrical Systems Engineering, Kyoto University of Advanced Science, Kyoto 615-8577, Japan}

\date{\today}

\begin{abstract}
The optical conductivity and the relevant electronic excitation processes are investigated in topologically-nontrivial MXenes, Mo$_2$HfC$_2$O$_2$ and W$_2$HfC$_2$O$_2$, utilizing first-principles calculation and effective model analysis.
The numerical calculation based on the first-principles band structure reveals the presence of several characteristic features in the spectrum of optical conductivity as a function of photon energy.
The drastic dependence on the photon polarization angle is also presented in terms of apparent features.
In this paper, an effective model is also generated referring to the crystal symmetries and applied to reveal the microscopic origin of the characteristics.
Then, it is shown that some features are strongly related to parity inversion between the conduction and valence bands, the key signature in electronic structures of topologically nontrivial insulators.
\end{abstract}

\maketitle
\section{Introduction}

%%%%%%%%%%%%%%%%%%%%%%%%%%%%%%%%%%%%%%%%%%%%%%

\begin{figure}[htbp]
\begin{center}
 \includegraphics[width=80mm]{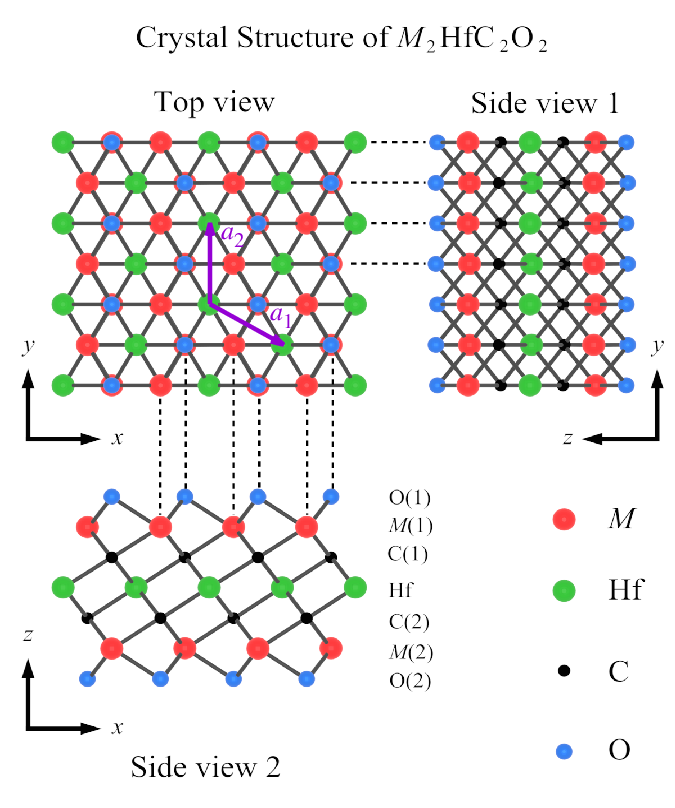}
\caption{The schematic overview of crystal structure for $M_2$HfC$_2$O$_2$ with $M=$ Mo or W. The $x$ and $y$ axes are adopted in two directions parallel to the layer, and the $z$ axis is perpendicular direction to the layer. Two lattice vectors are denoted by arrows, $\boldsymbol{a}_1$ and $\boldsymbol{a}_2$.
 }\label{fig_crystal_structure}
\end{center}
\end{figure}

MXenes are members of a large family of two-dimensional metal-carbides and nitrides, and have attracted much attention due to the wide range of physical and chemical properties exhibited by the two-dimensional materials via altering the chemical composition.\cite{Naguib2014,Anasori2015,Gogotsi2019}
Especially in some double transition-metal carbides and nitrides, topologically non-trivial electronic states have been theoretically predicted using first-principles calculations.\cite{Khazaei2016,Si2016,Liang2017,Huang2020,Dong2023,Parajuli2024}
Among the topologically nontrivial MXenes, the compound of Hf and group-VI transition metal, Mo$_2$HfC$_2$O$_2$ and W$_2$HfC$_2$O$_2$ (see Fig.\;\ref{fig_crystal_structure}), possesses a large energy gap and the stable surface terminated with O among the candidates.\cite{Khazaei2016,Si2016}
Since these MXenes possess time-reversal symmetric two-dimensional electronic states, they can exhibit helical edge channels leading to the quantum spin Hall effect as two-dimensional topological insulators.\cite{Kane2005,Bernevig2006}
Although the edge states are typical characteristics of topological insulators, it is not easy to confirm the time-reversal symmetric topological phase through the detection of the localized states.
This is because non-topological edge modes can emerge especially in the case of a two-dimensional material with complex structure, e.g., some sublayers or a multi-element compound, and they induce the Hall spin current which is not a conservative quantity and hard to detect accurately.

%%%%%%%%%%%%%%%%%%%%%%%%%%%%%%%%%%%%%%%%%%%%%%

%%%%%%%%%%%%%%%%%%%%%%%%%%%%%%%%%%%%%%%%%%%%%%

In the case of the topologically nontrivial MXenes, the crystal structure is complex due to the sublayers of several elements in comparison with conventional time-reversal two-dimensional topological insulators, e.g., the quantum well of CdTe and HgTe,\cite{Konig2007} $1T$'-transition metal ditelluride.\cite{Qian2014}
Thus, an alternative observable is used to confirm the topological nature in these two-dimensional materials.
In general, topological insulators also possess a characteristic electronic structure in the bulk, i.e., two opposite-parity bands and the energetic inversion of them around one or an odd number of high-symmetry wave numbers across the insulating bulk gap.
\cite{Bernevig2006-2,Fu2007,Murakami2007}
In Mo$_2$HfC$_2$O$_2$ and W$_2$HfC$_2$O$_2$,the parity inversion occurs between the conduction and valence bands only around the $\Gamma$ point according to the first-principles calculation.\cite{Khazaei2016}
Therefore, the confirmation of the parity inversion can be evidence of the topologically nontrivial phase of these materials.
In this paper, the optical conductivity of $M_2$HfC$_2$O$_2$ for $M=$Mo and W is investigated utilizing first-principles calculation and an effective model presented in Sec.\;\ref{sec_effective_model}.
The photon-energy dependence of optical conductivity provides information about not only the dispersion of electronic bands but also the wave function in the band, and thus, in practice, it has been investigated for several topologically nontrivial materials other than a topological insulator, Dirac or Weyl semimetal\cite{Xu2016,Neubauer2016,Tabert2016}, nodal-line semimetal,\cite{Schilling2017,Ahn2017,Barati2017,Habe2018}, multi-fold point-node semimetal,\cite{Sanchez2019,Habe2019,Xu2020}, and Hopf semimetal\cite{Habe2022}.
Thus, this paper shows the relation between the spectrum of the optical conductivity and the electronic structure associated with the topological states, parity inversion, in $M_2$HfC$_2$O$_2$ for $M=$Mo and W, theoretically through the effective model analysis.
Moreover, the characteristic photon polarization angel dependence is shown for the topologically nontrivial MXenes and the origin in the electronic states is also revealed through the effective model analysis.
The main body of this paper is organized as follows.
In Sec.\;\ref{sec_crystal_structure}, the crystal structure of $M_2$HfC$_2$O$_2$ for $M=$ Mo and W is presented performing the first-principles calculations.
In Sec.\;\ref{sec_electronic_structure}, the electronic structure is investigated in terms of the electronic excitation around the $\Gamma$ point analyzing the first-principles band structure.
Then, the optical conductivity of these MXenes is given utilizing a tight-binding model generated from the first-principles band structure in Sec.\;\ref{sec_optical_conductivity}.
Moreover, the joint density of states is numerically calculated for identifying the responsible bands for the characteristic feature in the spectrum of the optical conductivity.
In Sec.\;\ref{sec_effective_model}, an effective model describing electronic states around the $\Gamma$ point is derived for investigating the restrictions to the electronic excitation.
The analytic formula of a dynamical polarization matrix is derived and applied to the investigation of the relation between optical conductivity and electronic structure.
The discussion and conclusion are given in Sec.\;\ref{sec_conclusion}.

%%%%%%%%%%%%%%%%%%%%%%%%%%%%%%%%%%%%%%%%%%%%%%

\section{Crystal Structure}\label{sec_crystal_structure}

%%%%%%%%%%%%%%%%%%%%%%%%%%%%%%%%%%%%%%%%%%%%%%

The monolayer crystals of Mo$_2$HfC$_2$O$_2$ and W$_2$HfC$_2$O$_2$ are formed in the hexagonal lattice structure consisting of five sublayers of Hf, C, and Mo or W terminated by O atoms as shown in Fig.\;\ref{fig_crystal_structure}.
The hexagonal lattice structure is expanded by two lattice vectors, $\boldsymbol{a}_1=(\sqrt{3}a/2,-a/2)$ and $\boldsymbol{a}_2=(0,a)$ in the $xy$ plane with a single lattice constant $a$.
The thickness $c$ of the monolayer is defined by the distance between O atoms on the opposite surfaces along the $z$ axis. 
In table \ref{tab_atomic_position}, the position of each atom belonging to a unit cell is presented as the superposition of the basic lattice vectors and a perpendicular vector, $\boldsymbol{r}=x_1\boldsymbol{a}_1+x_2\boldsymbol{a}_2+x_3c\boldsymbol{e}_z$.
Here, the atomic coordinate in the $z$ direction can be represented by two parameters $d_{C-\mathrm{Hf}}$, the displacement along the $z$ axis between C and Hf, and $d_{M\mathrm{-Hf}}$, that between $M$ and Hf.

%%%%%%%%%%%%%%%%%%%%%%%%%%%%%%%%%%%%%%%%%%%%%%

\begin{table}
\caption{The atomic positions in a unit cell for $M_2$HfC$_2$O$_2$ for $M$=Mo and W. The coordinates $(x_1,x_2,x_3)$ are defined in the basis of two lattice vectors and a perpendicular vector, $\boldsymbol{r}=x_1\boldsymbol{a}_1+x_2\boldsymbol{a}_2+x_3c\boldsymbol{e}_z$.
}
\begin{ruledtabular}
\begin{tabular}{c c c c c c c c}
&O(1)&$M$(1)&C(1)&Hf&C(2)&$M$(2)&O(2)\\ \hline
$x_1$&2/3&1/3&2/3&0&-2/3&-1/3&-2/3\\ 
$x_2$&1/3&2/3&1/3&0&-1/3&-2/3&-1/3\\ 
$x_3$&1/2&$d_{M-\mathrm{Hf}}/c$&$d_{\mathrm{C-Hf}}/c$&0&$-d_{\mathrm{C-Hf}}/c$&$-d_{M\mathrm{-Hf}}/c$&-1/2\\ 
\end{tabular}\label{tab_atomic_position}
\end{ruledtabular}
\end{table}

%%%%%%%%%%%%%%%%%%%%%%%%%%%%%%%%%%%%%%%%%%%%%%

The material parameters, $a$, $c$, $d_{\mathrm{C-Hf}}$, and $d_{M\mathrm{-Hf}}$, for Mo$_2$HfC$_2$O$_2$ and W$_2$HfC$_2$O$_2$ are obtained numerically using the lattice optimization code, vc-relax in QUANTUM ESPRESSO,\cite{quantum-espresso} a package of numerical codes for density functional theory (DFT).
In the numerical calculation, the Predew-Burke-Emzerhof functional\cite{PBE_functional} in projector augmented wave method is adopted and the correction, vdw-df-C6,\cite{vdw-df-C6} to exchange correlation is applied for including the effect of van der Waals interaction.\cite{vdw-df,Sabatini2012}
The DFT calculation is performed on the $k$ mesh $12\times12\times1$ in the first Brillouin zone with the energy cutoff 60 Ry for the plane wave basis and 500 Ry for the charge density. 
For simulating the monolayer crystal, the lattice constant in the $z$ axis is fixed to 40 {\AA} for separating layers in the $z$ direction.
The convergence criterion is $10^{-8}$ Ry for self-consistent field calculation, $10^{-2}$ kbar for stress, and $10^{-4}$ Ry/Bohr for forces in lattice optimization calculations.
The numerically optimized parameters are presented in table \ref{tab_crystal_parameters}.

%%%%%%%%%%%%%%%%%%%%%%%%%%%%%%%%%%%%%%%%%%%%%%

\begin{table}
\caption{The numerically optimized material parameters of $M_2$HfC$_2$O$_2$ for $M$=Mo and W. These parameters are presented in the unit of {\AA}.
}
\begin{ruledtabular}
\begin{tabular}{c c c c c}
&$a$&$c$&$d_{M\mathrm{-Hf}}$&$d_{\mathrm{C-Hf}}$\\ \hline
Mo$_2$HfC$_2$O$_2$&2.990&7.670&2.655&1.378\\ 
W$_2$HfC$_2$O$_2$&2.990&7.749&2.685&1.408\\ 
\end{tabular}\label{tab_crystal_parameters}
\end{ruledtabular}
\end{table}

\begin{figure}[htbp]
\begin{center}
 \includegraphics[width=80mm]{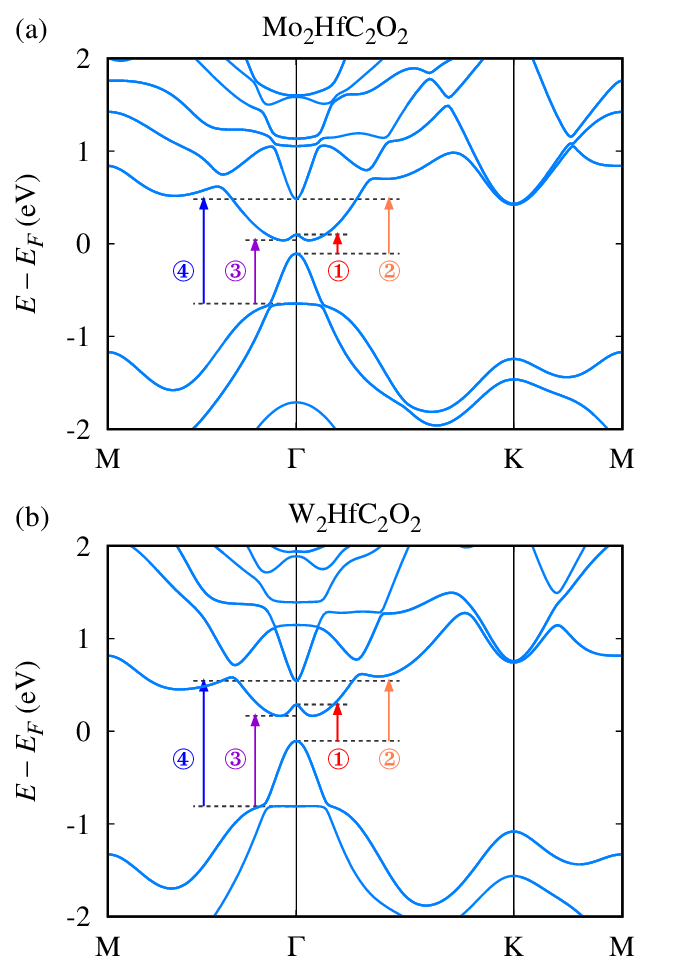}
\caption{The electronic band structures of (a) Mo$_2$HfC$_2$O$_2$ and (b) W$_2$HfC$_2$O$_2$ around the Fermi energy. The horizontal axis indicates the wave number along the path through high symmetry points M, $\Gamma$, and K. Each arrow represents an electronic excitation associated with the optical conductivity spectrum.
 }\label{fig_band_structure}
\end{center}
\end{figure}

%%%%%%%%%%%%%%%%%%%%%%%%%%%%%%%%%%%%%%%%%%%%%%

\section{Electronic Structure}\label{sec_electronic_structure}

%%%%%%%%%%%%%%%%%%%%%%%%%%%%%%%%%%%%%%%%%%%%%%

The electronic structure of $W_2$HfC$_2$O$_2$ is investigated applying the optimized material parameters to the first-principles band calculation.
The numerical calculation is also performed using QUANTUM ESPRESSO under the same condition for the lattice optimization.
In Fig.\;\ref{fig_band_structure}, the band structures in the presence of spin-orbit coupling are presented around the Fermi energy.
The conduction and valence bands split in the entire Brillouin zone, and these bands are close to each other around the $\Gamma$ point.
Thus, the optical response to low-energy photons is dominated by the electronic states around this high-symmetry point.
Moreover, according to the previous works,\cite{Khazaei2016,Si2016} the parity inversion between the conduction and valence bands, the key to the topological nature of the material, occurs at the $\Gamma$ point.
Therefore, the optical response to the low-energy photons can include the information about the electronic structure associated with the topological nature of these materials.

%%%%%%%%%%%%%%%%%%%%%%%%%%%%%%%%%%%%%%%%%%%%%%

In the spectrum of the optical response function, the electronic excitation around local minima or/and maxima in the band structure leads to characteristic enhancement at the photon frequency $\omega$ which coincides with the excitation energy.
Therefore, some characteristic excitation energies are schematically depicted in Fig.\;\ref{fig_band_structure}, and also presented in table \ref{tab_characteristic_energies}.
The list of excitation energies indicates the upper limit of photon energy $\hbar\omega<1.6$ eV associated with electronic states only around the $\Gamma$ point.
In this photon energy region, the four electronic bands, the two lowest conduction and two highest valence bands, are relevant to the optical properties of the topologically-nontrivial MXenes.
Since the electronic states on these bands are also responsible for the topological nature of these materials, the optical response under $\hbar\omega\simeq1.6$ eV is associated with the excitation among the topologically unconventional states.

%%%%%%%%%%%%%%%%%%%%%%%%%%%%%%%%%%%%%%%%%%%%%%

\begin{table}
\caption{The characteristic excitation energies in the band structures for Mo$_2$HfC$_2$O$_2$ and W$_2$HfC$_2$O$_2$. The excitation energies are presented in the unit of eV.
}
\begin{ruledtabular}
\begin{tabular}{c c c c c c c}
&$\Gamma(\textcircled{1})$&$\Gamma(\textcircled{2})$&$\Gamma(\textcircled{3})$&$\Gamma(\textcircled{4})$&K&M\\ \hline
Mo$_2$HfC$_2$O$_2$&0.206&0.689&0.745&1.128&1.664&2.010\\ 
W$_2$HfC$_2$O$_2$&0.394&0.648&0.973&1.352&1.823&2.142\\ 
\end{tabular}\label{tab_characteristic_energies}
\end{ruledtabular}
\end{table}

%%%%%%%%%%%%%%%%%%%%%%%%%%%%%%%%%%%%%%%%%%%%%%

\begin{figure}[htbp]
\begin{center}
 \includegraphics[width=80mm]{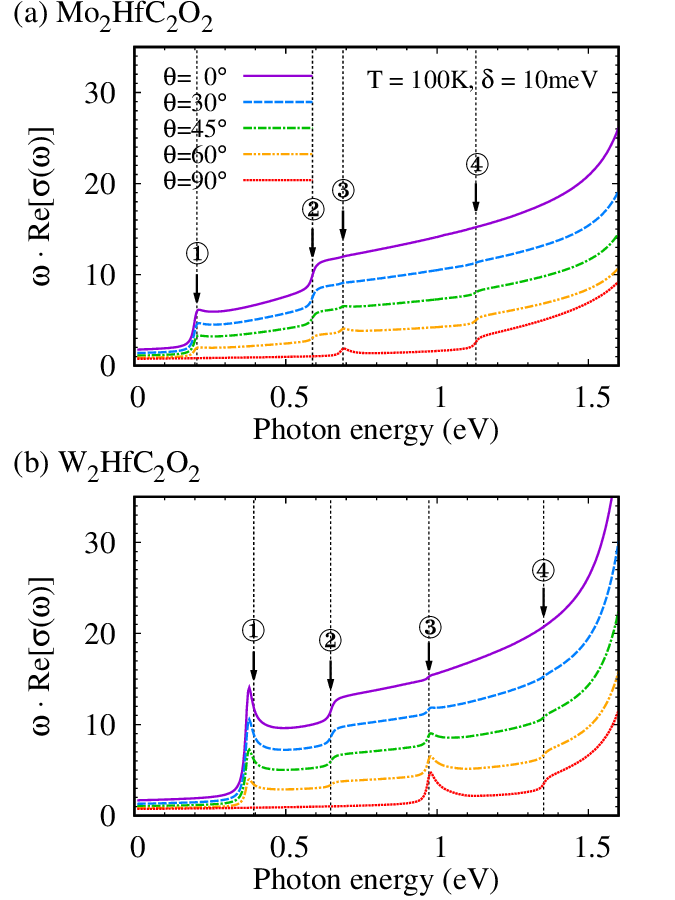}
\caption{The optical conductivity of (a) Mo$_2$HfC$_2$O$_2$ and (b) W$_2$HfC$_2$O$_2$ for the photon energy $\hbar\omega<1.6$ eV. The several curves represent the spectra for different angle of polarization with respect to the layer. ($\theta=0^\circ$ for the linear-polarized photon parallel to the $x$-axis and $\theta=90^\circ$ for that parallel to the $z$-axis)
 }\label{fig_dynamical_conductivity}
\end{center}
\end{figure}

%%%%%%%%%%%%%%%%%%%%%%%%%%%%%%%%%%%%%%%%%%%%%%

\section{Optical Conductivity}\label{sec_optical_conductivity}

%%%%%%%%%%%%%%%%%%%%%%%%%%%%%%%%%%%%%%%%%%%%%%

The optical properties of materials can be represented by the optical conductivity,
\begin{align}
\sigma(\omega,T)=&\frac{i}{\omega}\frac{e^2}{\hbar}\sum_{m\leq n}\int_{\mathrm{BZ}}\frac{d^2\boldsymbol{k}}{(2\pi)^2}\frac{\left|\langle n \boldsymbol{k}|\hat{\dot{r}}_{\theta,\phi}|m\boldsymbol{k}\rangle\right|^2}{\omega-(\omega_{n\boldsymbol{k}}-\omega_{m\boldsymbol{k}})+i\delta/\hbar}\nonumber\\
&\times\left(n_F(E_{n\boldsymbol{k}},T)-n_F(E_{m\boldsymbol{k}},T)\right),\label{eq_dynamical_conductivity}
\end{align}
which describes the dynamical response of electronic system to linearly-polarized light with the frequency $\omega$.
The dynamical response includes all direct excitation processes among the electronic bands $E_{m\boldsymbol{k}}$ in the Brillouin zone (BZ) where the occupancy of each state is represented by Fermi distribution $n_F(E,T)$ at the temperature $T$.
The self-energy is approximated by a constant relaxation time $\tau=\hbar/(2\delta)$.
Here, the time-derivative of position operator, $\hat{\dot{r}}_{\theta,\phi}=(1/i\hbar)[\hat{r}_{\theta,\phi},\hat{H}]$, represents the dynamical electronic polarization along the parallel axis to the photon polarization direction denoted by $(\theta,\phi)$ with $\hat{r}_{\theta,\phi}=\hat{x}\sin\theta\cos\phi +\hat{y}\sin\theta\sin\phi+\hat{z}\cos\theta$.
The Hamiltonian $\hat{H}$, the state vector $|m\boldsymbol{k}\rangle$, and the energy $E_{m\boldsymbol{k}}$ are obtained utilizing a multi-orbital tight-binding model.
The operators can be obtained as the velocity operators, $\hat{\dot{x}}=(1/i\hbar)(d\hat{H}_{\mathrm{tot}}/dk_x)$ and $\hat{\dot{y}}=(1/i\hbar)(d\hat{H}_{\mathrm{tot}}/dk_y)$, for the in-plane coordinates, and that along the $z$ direction is obtained using the original definition with the origin at the middle sublayer of Hf atoms.
The hopping parameters for the tight-binding model are calculated using Wannier90,\cite{wannier90} a numerical code to generate the maximally-localized Wannnier functions and the hopping integrals among them well reproducing the first principles band structure.
In this calculation, three $p$-orbitals in O and C, and five $d$-orbitals in Mo, W, and Hf are adopted as Wannier orbitals of the basis.

%%%%%%%%%%%%%%%%%%%%%%%%%%%%%%%%%%%%%%%%%%%%%%

The optical conductivity of Mo$_2$HfC$_2$O$_2$ and W$_2$HfC$_2$O$_2$ is numerically obtained for several polarization angles of photon polarization $\theta$ from the perpendicular direction, the $z$ axis, to the layer, and its real part is presented in Fig.\;\ref{fig_dynamical_conductivity}.
The calculations are performed under the fixed temperature T=100K and self-energy $\delta=10$meV.
%The incident photon is assumed to be linearly-polarized in the $xz$ plane. 
In the spectrum, there are four features characterizing the materials, two peaks and two step-like increments, up to 1.6 eV in Fig.\;\ref{fig_dynamical_conductivity}.
For both compounds of Mo and W, one peak and one step are hidden for $\theta=0^\circ$ and $90^\circ$, the perpendicular polarization and the parallel polarization to the layer, respectively.
The photon energies to induce these specific features seem to correspond to the excitation energies indicated in Fig.\;\ref{fig_band_structure}, although there is a small deviation in the case of \textcircled{1}, which is discussed in the following section.

%%%%%%%%%%%%%%%%%%%%%%%%%%%%%%%%%%%%%%%%%%%%%%

\begin{figure}[htbp]
\begin{center}
 \includegraphics[width=80mm]{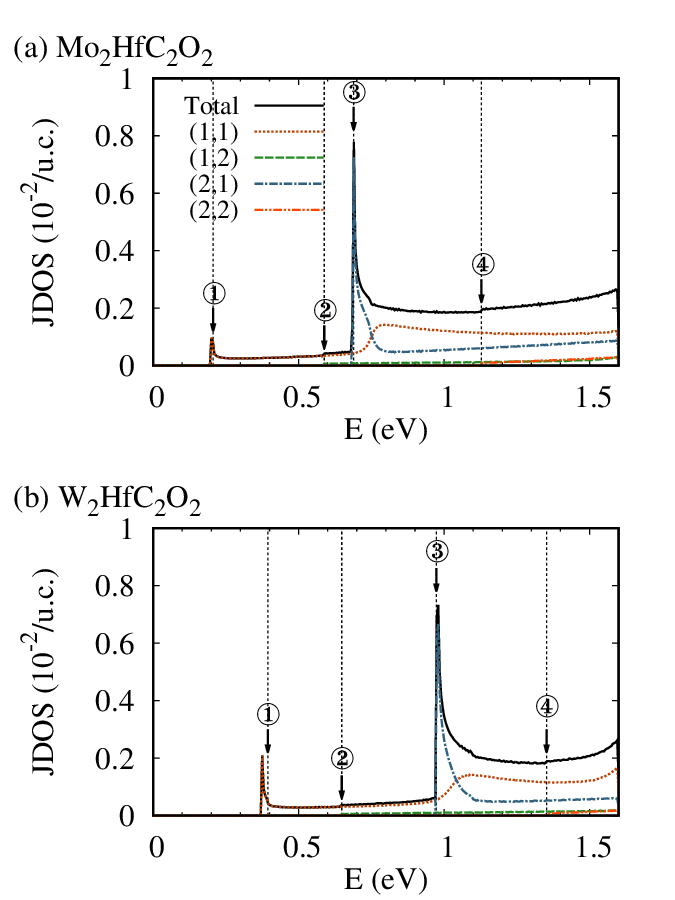}
\caption{The joint density of states among the two highest valence bands and the two lowest conduction bands for (a) Mo$_2$HfC$_2$O$_2$ and (b) W$_2$HfC$_2$O$_2$. The vertical dashed lines indicate the excitation energies represented in Fig.\;\ref{fig_band_structure}.
 }\label{fig_JDOS}
\end{center}
\end{figure}

%%%%%%%%%%%%%%%%%
%%%%%%%%%%%%%%%%%%%%%%%%%%%%%

The relation between the spectrum of $\sigma(\omega)$ and the band structure is conventionally represented by the joint density of states (JDOS).
The JDOS $J_{mn}(E)$ between a valence band $m$ and a conduction band $n$ is defined as follows,
\begin{align}
J_{mn}(E)&=\int_{BZ}\frac{d^2\boldsymbol{k}}{(2\pi)^2}\frac{1}{\pi}\mathrm{Re}\left[\lim_{\delta\rightarrow0}\frac{i}{E-(E_{n\boldsymbol{k}}-E_{m\boldsymbol{k}})+i\delta}\right]\nonumber\\
&=\int_{\mathrm{BZ}}\frac{d^2\boldsymbol{k}}{(2\pi)^2}\delta(E_{n\boldsymbol{k}}-E_{m\boldsymbol{k}}-E),\label{eq_JDOS}
\end{align}
and it gives the density of pairs of electronic states in the conduction and valence bands with the specific energy difference $E$.
The first formula describes that the JDOS represents the contribution of the denominator in Eq.\;(\ref{eq_dynamical_conductivity}) and the second one is utilized for actual calculations.
The numerically obtained JDOS is presented in Fig.\;\ref{fig_JDOS}(a) and (b) for Mo$_2$HfC$_2$O$_2$ and W$_2$HfC$_2$O$_2$, respectively.
Here, the pair of indexes $(i,j)$ indicates the combination of conduction and valence bands, where the first index $i$ denotes the top and second top valence bands for $i=1$ and 2, respectively, and the second index $j$ denotes the bottom and second bottom conduction bands for $j=1$ and $2$, respectively. 

%%%%%%%%%%%%%%%%%%%%%%%%%%%%%%%%%%%%%%%%%%%%%%

The JDOS shows that the peak-like enhancement of $\sigma(\omega,T)$, \textcircled{1} and \textcircled{2}, occurs at the energy for the peak of $J_{m1}(E)$ associated with the electronic excitation to the lowest conduction band.
On the other hand, the step-like increments correspond to $J_{m2}(E)$ associated with the excitation to the second lowest conduction band.
The numerical calculation reveals that the excitation to the second lowest conduction band gives a relatively large contribution to the optical conductivity in spite of much smaller JDOS compared with the excitation to the lowest conduction band.
Therefore, the enhancement via the excitation to the second lowest conduction band and the hidden peak depending on the polarization angle $\theta$ are attributed to the velocity component, the numerator, in Eq.\;(\ref{eq_dynamical_conductivity}).

%%%%%%%%%%%%%%%%%%%%%%%%%%%%%%%%%%%%%%%%%%%%%%

\begin{figure}[htbp]
\begin{center}
 \includegraphics[width=80mm]{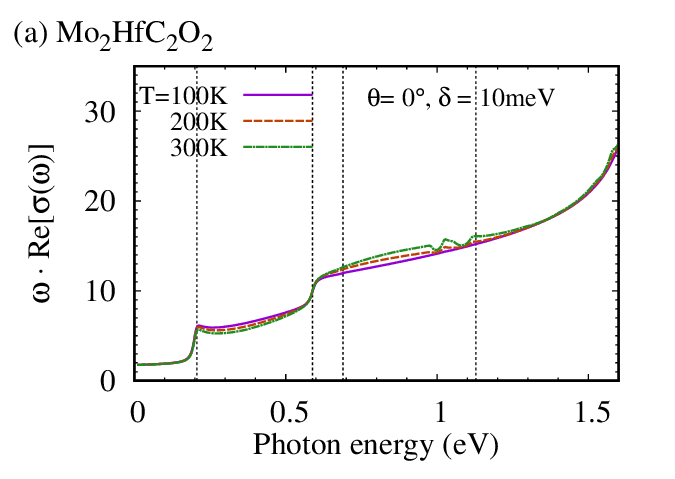}
\caption{The optical conductivity of Mo$_2$HfC$_2$O$_2$ for several temperatures, $T=100$K, 200K, and 300K. The polarization angle $\theta$ and self-energy $\delta$ are $0^\circ$ and $10$meV, respectively. 
 }\label{fig_dynamical_conductivity_temp}
\end{center}
\end{figure}

%%%%%%%%%%%%%%%%%%%%%%%%%%%%%%%%%%%%%%%%%%%%%%

Finally, the effect of thermally fluctuation is discussed in terms of the optical conductivity. 
At non-zero temperature, electrons are thermally excited to the conduction band and partially occupy the electronic states on the band.
In this case, the partially occupied states on the conduction band can be also initial states for optical excitation.
In Fig. \;\ref{fig_dynamical_conductivity_temp},the optical conductivity of Mo$_2$HfC$_2$O$_2$ is presented for several temperatures, $T=100$K, 200K, and 300K.
At $T=300$K, an additional small peak appears just above $\hbar\omega=1$eV, and it is attributed to the thermal excited electrons in the lowest conduction band.
The temperature dependence cannot be observed in the case of W$_2$HfC$_2$O$_2$ because of the larger insulating gap in comparison with Mo$_2$HfC$_2$O$_2$.
Although the numerical calculation clearly indicates the thermal effect in the case of Mo$_2$HfC$_2$O$_2$, this effect is not ensured in the actual experiments because the band gap can be underestimated in DFT.
Thus, the low-temperature spectrum in Fig.\;\ref{fig_dynamical_conductivity} can be observed even at room temperature with an energy shift due to the increase of the band gap.

%%%%%%%%%%%%%%%%%%%%%%%%%%%%%%%%%%%%%%%%%%%%%%

\section{Effective model analysis}\label{sec_effective_model}

%%%%%%%%%%%%%%%%%%%%%%%%%%%%%%%%%%%%%%%%%%%%%%

\subsection{Effective Hamiltonian without SOC}

%%%%%%%%%%%%%%%%%%%%%%%%%%%%%%%%%%%%%%%%%%%%%%

In this section, the polarization-angle dependence of electronic excitation process are investigated using an effective model to describe the electronic states around the $\Gamma$ point for Mo$_2$HfC$_2$O$_2$ and W$_2$HfC$_2$O$_2$.
As shown in Eq.\;(\ref{eq_dynamical_conductivity}), the properties are represented by the dynamical polarization matrix $\langle n\boldsymbol{k}|\hat{\dot{r}}_{\theta,\phi}|m\boldsymbol{k}\rangle$.
Firstly, the effective model is generated for the two conduction bands and two valence bands in the absence of SOC.
At the $\Gamma$ point, the electronic states are invariant under some symmetrical operations, spatial inversion, threefold rotation, and mirror operation along the $y$ axis, because of the crystal symmetry.
Since threefold rotation and mirror operation do not commute, spatial parity $P=\pm$ and the eigenvalue for threefold rotation $R_3=1$, $\Omega$, and $\Omega^\ast$ with $\Omega=\exp[2\pi i/3]$ are adopted for characterizing electronic states $|R_3,P\rangle$. 
In the absence of SOC, two bands with $(\Omega,P)$ and $(\Omega^\ast,P)$ must be degenerate at the high-symmetry point because the mirror operation interchanges them.
On the other hand, the bands with $R_3=1$ can be isolated from the other bands.

%%%%%%%%%%%%%%%%%%%%%%%%%%%%%%%%%%%%%%%%%%%%%%

\begin{figure}[htbp]
\begin{center}
 \includegraphics[width=80mm]{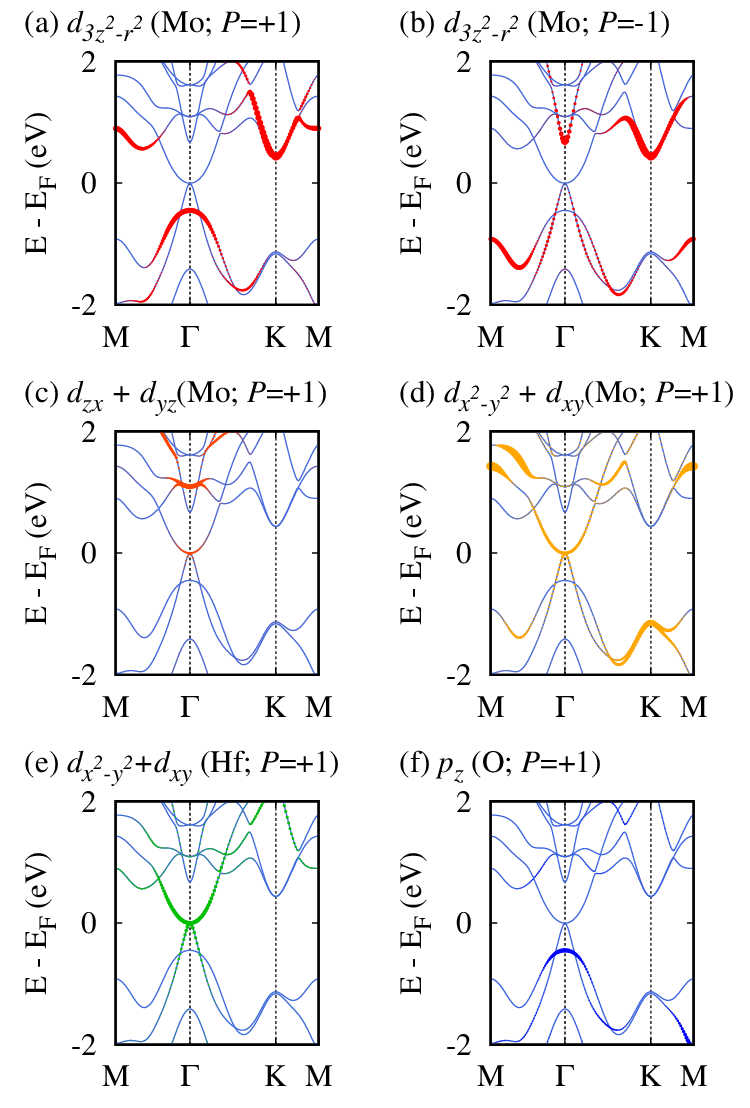}
\caption{
The band structure without SOC and the wave amplitude of each Wannier orbital in electronic states. The amplitude of each Wannier function is represented by the symbol size on the band structure of Mo$_2$HfC$_2$O$_2$.
 }\label{fig_orbital_amplitude}
\end{center}
\end{figure}

To identify the character of actual bands, the amplitude of each Wannier orbital in the electronic states is analyzed utilizing the multi-orbital tight-binding model generated by Wannier90.
However, the Wannier orbitals in the basis are slightly modified from the orbitals localized at each atomic position.
Since Mo, W, C, and O, are not placed at the inversion center of the unit cell as shown in table \ref{tab_atomic_position}, the Wannier orbitals in these atoms are not inversion symmetric.
However, the superposition of the same orbital $w$ in an element $A$,
\begin{align}
\frac{1}{\sqrt{2}}(|w, A(1)\rangle\pm|w,A(2)\rangle),
\end{align}
can be a Wanneir function preserving spatial parity if $w$ is an orbital characterized by a magnetic quantum number.
The spacial parity $P$ of the renewed Wannier function depends on the parity of the orbital $w$, i.e., odd parity for $p$-orbitals and even parity for $d$-orbitals.
Moreover, it is also affected by the relative sign in two atomic positions $A(1)$ and $A(2)$.
Then, the even parity function $|w,+\rangle$ and the odd parity functions $|w,-\rangle$ can be obtained as follows,
\begin{align}
|p_\alpha,\pm\rangle=\frac{1}{\sqrt{2}}(|p_\alpha, A(1)\rangle\mp|p_\alpha,A(2)\rangle),\\
|d_\alpha,\pm\rangle=\frac{1}{\sqrt{2}}(|d_\alpha, A(1)\rangle\pm|d_\alpha,A(2)\rangle).
\end{align}
In Fig.\;\ref{fig_orbital_amplitude},the amplitude of relevant orbitals to the bands without SOC is presented on the band structure of  Mo$_2$HfC$_2$O$_2$ around the Fermi energy. 
The conduction and valence bands touch the Fermi energy at the $\Gamma$ point. 
The degenerate states consist of even parity orbitals with non-zero angular momentum along the $z$ axis, i.e., $d_{zx}$, $d_{yz}$, $d_{x^2-y^2}$, and $d_{xy}$.
On the other hand, zero angular momentum orbitals, $d_{3z^2-r^2}$ and $p_z$, appear on the second highest valence band and the second lowest conduction band with even and odd parities, respectively. 
Therefore, at the $\Gamma$ point, the four electronic states are characterized by two indexes $(R_3,P)$ as follows,
\begin{align}
(|1,-\rangle, |\Omega,+\rangle, |\Omega^*,+\rangle, |1,+\rangle),\label{eq_basis}
\end{align}
in order of energy.
Then, the effective Hamiltonian for $\boldsymbol{k}=0$ is a diagonal matrix on this basis, 
\begin{align}
\hat{H}_G=
\mathrm{diag}[E_{t},0,0,E_{b}],
\end{align}
where the Fermi energy is set to zero.
The energy at the band edge for the top and bottom bands is represented by $E_t$ and $E_b$, respectively.

%%%%%%%%%%%%%%%%%%%%%%%%%%%%%%%%%%%%%%%%%%%%%%

To describe electronic states around the $\Gamma$ point, the additional terms $\hat{H}(k,\varphi_{\boldsymbol{k}})$ to $\hat{H}_G$ are considered up to the second order of wave number $k^2$, where $\varphi_{\boldsymbol{k}}$ describes the direction of the wave number as $\tan\varphi_{\boldsymbol{k}}=k_y/k_x$. 
The crystal structure imposes some restrictions on $\hat{H}(k,\varphi_{\boldsymbol{k}})$ due to the invariance under the symmetrical operations, spatial inversion
$\hat{P}$, threefold rotation $\hat{R}_3$, and mirror operation along the $y$-axis, $\hat{M}_y$,
\begin{align}
\hat{P}^\dagger \hat{H}(k,\varphi_{\boldsymbol{k}})\hat{P}=&U_P^\dagger \hat{H}(k,\varphi_{\boldsymbol{k}}-\pi)U_P,\nonumber\\
\hat{R}_3^\dagger \hat{H}(k,\varphi_{\boldsymbol{k}})\hat{R}_3=&U_{R_3}^\dagger \hat{H}(k,\varphi_{\boldsymbol{k}}-2\pi/3)U_{R_3},\\
\hat{M}_y^\dagger \hat{H}(k,\varphi_{\boldsymbol{k}})\hat{M}_y=&U_{M_y}^\dagger \hat{H}(k,-\varphi_{\boldsymbol{k}})U_{M_y},\nonumber
\end{align}
with unitary matrices, $U_P=\mathrm{diag}[-1,1,1,1]$, $U_{R_3}=\mathrm{diag}[1,e^{2\pi i/3},e^{-2\pi i/3},1]$, and 
\begin{align}
U_{M_y}=\begin{pmatrix}
1&0&0&0\\
0&0&1&0\\
0&1&0&0\\
0&0&0&1
\end{pmatrix}.
\end{align}
These conditions derive the following form for $\hat{H}(k,\theta_{\boldsymbol{k}})$,
\begin{align}
\hat{H}(k,\varphi_{\boldsymbol{k}})=\begin{pmatrix}
\beta_2k^2&\alpha ke^{i\varphi_{\boldsymbol{k}}}&\alpha ke^{-i\varphi_{\boldsymbol{k}}}&0\\
\alpha ke^{-i\varphi_{\boldsymbol{k}}}&\beta_1k^2&-\gamma k^2e^{-2i\varphi_{\boldsymbol{k}}}&0\\
\alpha ke^{i\varphi_{\boldsymbol{k}}}&-\gamma k^2e^{2i\varphi_{\boldsymbol{k}}}&\beta_1k^2&0\\
0&0&0&\beta_2k^2
\end{pmatrix},\label{eq_k-dependence}
\end{align}
where the mixing between $|1,+\rangle$ and others is omitted due to the numerical result in Fig.\;\ref{fig_orbital_amplitude}.
Here, the diagonal components for $|1,\pm\rangle$ are assumed to be equivalent because the Wannier functions consist of the same orbital, $d_{3z^2-r^2}$, mainly and the intra-sublayer orbital-hybridization is relevant for these components. 

%%%%%%%%%%%%%%%%%%%%%%%%%%%%%%%%%%%%%%%%%%%%%%

\subsection{Effective Hamiltonian and Polarization Operator with SOC}

%%%%%%%%%%%%%%%%%%%%%%%%%%%%%%%%%%%%%%%%%%%%%%

In the presence of threefold rotation symmetry, the SOC is represented by $\hat{H}_{\mathrm{SOC}}=\lambda\hat{l}_z\cdot\hat{s}_z$ with angular momentum operators $\hat{l}_z$ and $\hat{s}_z$ for the orbital and the spin, respectively, along the perpendicular direction to the layer.
Since the orbital angular momentum is up to 2 for the $p$- and $d$-orbitals, the state $|\pm,1\rangle$ consists of orbitals with $l_z=0$, and $|+,\Omega\rangle$ and $|+,\Omega^\ast\rangle$ mainly consist of orbitals with the opposite orbital angular momenta, e.g., $l_z=-2$ and $2$.
Thus, the SOC can be introduced as follows,
\begin{align}
 \hat{H}_{\mathrm{SOC}}=\mathrm{diag}[0,-\lambda \sigma_z,\lambda \sigma_z,0],
\end{align}
for spin $\sigma_z=2s_z=\pm1$ on the basis in Eq.\;(\ref{eq_basis}).
Therefore, the total effective Hamiltonian around the $\Gamma$ point is represented by
\begin{align}
\hat{H}_{\mathrm{tot}}=\hat{H}_G+\hat{H}(k,\theta_{\boldsymbol{k}})+\hat{H}_{\mathrm{SOC}}.\label{eq_effective_Hamiltonian}
\end{align}

%%%%%%%%%%%%%%%%%%%%%%%%%%%%%%%%%%%%%%%%%%%%%%

To analyze the off-diagonal component of dynamical polarization matrix in Eq.\;(\ref{eq_dynamical_conductivity}), their operators are derived within the effective model in Eq.\;(\ref{eq_effective_Hamiltonian}).
The in-plane component is equivalent to the velocity operator and thus it can be defined as the derivative of the Hamiltonian, $\hat{\dot{r}}_\mu=\hat{v}_\mu=(1/i\hbar)(d\hat{H}_{\mathrm{tot}}/dk_\mu)$, given by
\begin{align}
\hat{\dot{x}}=&\frac{1}{\hbar}\begin{pmatrix}
2\beta_2k_x&\alpha&\alpha&0\\
\alpha&2\beta_1k_x&-2\gamma ke^{-i\varphi_{\boldsymbol{k}}}&0\\
\alpha&-2\gamma ke^{i\varphi_{\boldsymbol{k}}}&2\beta_1k_x&0\\
0&0&0&2\beta_2k_x
\end{pmatrix},\\
\hat{\dot{y}}=&\frac{1}{\hbar}\begin{pmatrix}
2\beta_2k_y&i\alpha&-i\alpha&0\\
-i\alpha&2\beta_1k_y&2i\gamma ke^{-i\varphi_{\boldsymbol{k}}}&0\\
i\alpha&-2i\gamma ke^{i\varphi_{\boldsymbol{k}}}&2\beta_1k_y&0\\
0&0&0&2\beta_2k_y
\end{pmatrix}.
\end{align}
The perpendicular component to the layer is defined in a different manner, i.e., use of the Heisenberg form $\hat{\dot{z}}=(1/i\hbar)[\hat{z},\hat{H}_{\mathrm{tot}}]$ directly with the position operator $\hat{z}$, due to the lack of the periodicity.
Since the position operator $\hat{z}$ changes its sign under spatial inversion, the off-diagonal components are non-zero only between two states with opposite parities.
Moreover, the invariance of $\hat{z}$ under threefold rotation requires the two states possessing the same eigenvalue under this symmetrical operation $\hat{R_3}$.
Therefore, the position operator $\hat{z}$ has a non-zero component only between $|1,+\rangle$ and $|1,-\rangle$ on the basis,
\begin{align}
\hat{z}=\begin{pmatrix}
0&0&0&z_0\\
0&0&0&0\\
0&0&0&0\\
z_0&0&0&0
\end{pmatrix},
\end{align}
where the origin of the $z$-axis is set to the center of the layer.
Thus, the time-derivative of $\hat{z}$ can be represented by
\begin{align}
\hat{\dot{z}}=\frac{z_0}{i\hbar}\begin{pmatrix}
0&0&0&-\Delta E\\
0&0&0&-\alpha ke^{-i\varphi_{\boldsymbol{k}}}\\
0&0&0&-\alpha ke^{i\varphi_{\boldsymbol{k}}}\\
\Delta E&\alpha ke^{i\varphi_{\boldsymbol{k}}}&\alpha ke^{-i\varphi_{\boldsymbol{k}}}&0
\end{pmatrix},
\end{align}
with the energy difference $\Delta E=E_t-E_b$.\\

%%%%%%%%%%%%%%%%%%%%%%%%%%%%%%%%%%%%%%%%%%%%%%

\subsection{A Simplified Case $\beta_j=0$ and $\alpha k\ll E_t$}

%%%%%%%%%%%%%%%%%%%%%%%%%%%%%%%%%%%%%%%%%%%%%%

For analyzing the characteristic features in Fig.\;\ref{fig_dynamical_conductivity}, an approximation $\beta_j=0$ is introduced to simplify the effective model in Eq.\;(\ref{eq_effective_Hamiltonian}).
Even under the condition, the model does not lose the nature about the mixing of orbitals in electronic states around the $\Gamma$ point.
Without the SOC term, i.e., $\hat{H}_G+\hat{H}(k,\theta_{\boldsymbol{k}})$, the simplified model gives two conduction bands,
\begin{align}
\begin{split}
E_2^{(0)}=&E_t+2\Delta_k,\\
E_1^{(0)}=&\gamma k^2,\\
\end{split}
\end{align}
and two valence bands,
\begin{align}
\begin{split}
E_{-1}^{(0)}=&-\gamma k^2-2\Delta_k,\\
E_{-2}^{(0)}=&E_b,
\end{split}
\end{align}
where $\Delta_k$ represents the dispersion of the second lowest conduction band $E_2^{(0)}$,
\begin{align}
\Delta_k=\frac{1}{4}\sqrt{(E_t+\gamma k^2)^2+8\alpha^2k^2}-\frac{E_t+\gamma k^2}{4}.
\end{align}
The eigenstate vectors corresponding to these bands are given by
\begin{align}
\begin{split}
\psi_2^{(0)}=&\frac{1}{2}\begin{pmatrix}
\sqrt{2(1+\cos\phi_{\boldsymbol{k}})}\\
e^{-i\varphi_{\boldsymbol{k}}}\sqrt{1-\cos\phi_{\boldsymbol{k}}}\\
e^{i\varphi_{\boldsymbol{k}}}\sqrt{1-\cos\phi_{\boldsymbol{k}}}\\
0
\end{pmatrix},\;\;
\psi_1^{(0)}=\frac{1}{\sqrt{2}}\begin{pmatrix}
0\\
-e^{-i\varphi_{\boldsymbol{k}}}\\
e^{i\varphi_{\boldsymbol{k}}}\\
0
\end{pmatrix},\\
\psi_{-1}^{(0)}=&\frac{1}{2}\begin{pmatrix}
-\sqrt{2(1-\cos\phi_{\boldsymbol{k}})}\\
e^{-i\varphi_{\boldsymbol{k}}}\sqrt{1+\cos\phi_{\boldsymbol{k}}}\\
e^{i\varphi_{\boldsymbol{k}}}\sqrt{1+\cos\phi_{\boldsymbol{k}}}\\
0
\end{pmatrix},\;\;
\psi_{-2}^{(0)}=\frac{1}{\sqrt{2}}\begin{pmatrix}
0\\
0\\
0\\
1
\end{pmatrix},
\end{split}
\end{align}
with the phase factor $\phi_{\boldsymbol{k}}$ defined as
\begin{align}
\cos\phi_{\boldsymbol{k}}=(E_t+\gamma k^2)/\sqrt{(E_t+\gamma k^2)^2+8\alpha^2 k^2}.
\end{align}
Since $\cos\phi_{\boldsymbol{k}}$ represents the $\boldsymbol{k}$ dependent mixing between the odd parity state $|1,-\rangle$ and even parity states $(|\Omega,+\rangle,\;|\Omega^\ast,+\rangle)$, the component $\alpha ke^{\pm i\varphi_{\boldsymbol{k}}}$ is relevant to the parity inversion between the conduction band and the valence band, i.e., the key hybridization to topologically unconventional properties of Mo$_2$HfC$_2$O$_2$ and W$_2$HfC$_2$O$_2$.

%%%%%%%%%%%%%%%%%%%%%%%%%%%%%%%%%%%%%%%%%%%%%%

In the presence of the SOC term, it is useful to represent the total Hamiltonian on the basis of $(\psi^{(0)}_2,\psi^{(0)}_1,\psi^{(0)}_{-1},\psi^{(0)}_{-2})$ as follows,
\begin{align}
\hat{H}_{\mathrm{tot}}=\begin{pmatrix}
E_2^{(0)}&\lambda{\sigma_z}\sin\frac{\phi_{\boldsymbol{k}}}{2}&0&0\\
\lambda{\sigma_z}\sin\frac{\phi_{\boldsymbol{k}}}{2}&E_1^{(0)}&\lambda{\sigma_z}\cos\frac{\phi_{\boldsymbol{k}}}{2}&0\\
0&\lambda{\sigma_z}\cos\frac{\phi_{\boldsymbol{k}}}{2}&E_{-1}^{(0)}&0\\
0&0&0&E_{-2}^{(0)}\\
\end{pmatrix}.
\end{align}
In the vicinity of the $\Gamma$ point, the SOC between the two conduction bands, $\psi_2$ and $\psi_1$, is negligibly small because of $2\sqrt{2}\alpha k/E_t\ll 1$, i.e., $\cos\phi_{\boldsymbol{k}}\simeq1$. 
Then, in this range of $k$, the Hamiltonian contains only the coupling between the lowest conduction and the highest valence bands,
\begin{align}
\hat{H}_{\mathrm{tot}}\simeq\begin{pmatrix}
E_-+2\Delta_k&0&0&0\\
0&\gamma k^2&\lambda{\sigma_z}&0\\
0&\lambda{\sigma_z}&-\gamma k^2-2\Delta_k&0\\
0&0&0&E_+\\
\end{pmatrix}.
\end{align}
Thus the second lowest conduction band $E^{(0)}_2$ and the second highest valence band $E_{-2}^{(0)}$ remain unchanged even in the presence of SOC term,
\begin{align}
\begin{split}
E_{\pm2}^{\sigma_z}=E^{(0)}_{\pm2}.
\end{split}
\end{align}
On the other hand, the other two bands $E_{\pm1}^{(0)}$ near the Fermi energy are hybridized and split as
\begin{align}
\begin{split}
E_{\pm1}^{\sigma_z}=&\pm\lambda_k'-\Delta_k,
\end{split}
\end{align}
with 
\begin{align}
\lambda_k'=\sqrt{\lambda^2+(\gamma k^2+\Delta_k)^2}.
\end{align}
Since the crystals possess both time-reversal and spatial inversion symmetries, all the bands are doubly degenerate due to Kramer's degeneracy.
However, the eigenstate vectors with SOC explicitly depend on the spin direction $\sigma_z$ as follows,
\begin{align}
\begin{split}
\psi_2^{\sigma_z}=&\begin{pmatrix}
1\\
0\\
0\\
0
\end{pmatrix},\\
\psi_1^{\sigma_z}=&\frac{\sqrt{1+\eta_{\boldsymbol{k}}}}{2}\begin{pmatrix}
0\\
e^{-i\varphi_{\boldsymbol{k}}}\\
-e^{i\varphi_{\boldsymbol{k}}}\\
0
\end{pmatrix}
+\frac{\sigma_z\sqrt{1-\eta_{\boldsymbol{k}}}}{2}
\begin{pmatrix}
0\\
e^{-i\varphi_{\boldsymbol{k}}}\\
e^{i\varphi_{\boldsymbol{k}}}\\
0
\end{pmatrix}
,
\\
\psi_{-1}^{\sigma_z}=&\frac{\sigma_z\sqrt{1-\eta_{\boldsymbol{k}}}}{2}\begin{pmatrix}
0\\
e^{-i\varphi_{\boldsymbol{k}}}\\
-e^{i\varphi_{\boldsymbol{k}}}\\
0
\end{pmatrix}
-
\frac{\sqrt{1+\eta_{\boldsymbol{k}}}}{2}\begin{pmatrix}
0\\
e^{-i\varphi_{\boldsymbol{k}}}\\
e^{i\varphi_{\boldsymbol{k}}}\\
0
\end{pmatrix},\\
\psi_{-2}^{\sigma_z}=&\begin{pmatrix}
0\\
0\\
0\\
1
\end{pmatrix},
\end{split}
\end{align}
with $\eta_{\boldsymbol{k}}=\tilde{\gamma}k^2/\sqrt{\tilde{\gamma}^2k^4+\lambda^2}$.
Therefore, the off-diagonal components of dynamical polarization matrix relevant to the excitation from the highest valence band $E_{-1}$ are non-zero except for the $z$ component as follows,
\begin{align}
\begin{split}
\langle\psi^{\sigma_z}_2|\hat{\dot{x}}|\psi^{\sigma_z}_{-1}\rangle=&-\frac{\alpha}{\hbar}(\sqrt{1+\eta_{\boldsymbol{k}}}\cos\varphi_{\boldsymbol{k}}+i\sigma_z \sqrt{1-\eta_{\boldsymbol{k}}}\sin\varphi_{\boldsymbol{k}}),\\
\langle\psi^{\sigma_z}_1|\hat{\dot{x}}|\psi^{\sigma_z}_{-1}\rangle=&\frac{2\gamma}{\hbar}k\left(\sqrt{1-\eta_{\boldsymbol{k}}^2}\cos\varphi_{\boldsymbol{k}}+i\sin\varphi_{\boldsymbol{k}}\right),\\
\langle\psi^{\sigma_z}_2|\hat{\dot{y}}|\psi^{\sigma_z}_{-1}\rangle=&-\frac{\alpha}{\hbar}(\sqrt{1+\eta_{\boldsymbol{k}}}\sin\varphi_{\boldsymbol{k}}-i\sigma_z\sqrt{1-\eta_{\boldsymbol{k}}}\cos\varphi_{\boldsymbol{k}}),\\
\langle\psi^{\sigma_z}_1|\hat{\dot{y}}|\psi^{\sigma_z}_{-1}\rangle=&\frac{2\gamma}{\hbar}k\left(\sqrt{1-\eta_{\boldsymbol{k}}^2}\sin\varphi_{\boldsymbol{k}}-i\cos\varphi_{\boldsymbol{k}}\right),\\
\langle\psi^{\sigma_z}_2|\hat{\dot{z}}|\psi^{\sigma_z}_{-1}\rangle=&\langle\psi^{\sigma_z}_1|\hat{\dot{z}}|\psi^{\sigma_z}_{-1}\rangle=0,\label{eq_excitation_1}
\end{split}
\end{align}
but those relevant to the excitation from the second highest valence band $E_{-2}$ remain non-zero only in the case of the $z$ component,
\begin{align}
\begin{split}
\langle\psi^{\sigma_z}_2|\hat{\dot{x}}|\psi^{\sigma_z}_{-2}\rangle=&\langle\psi^{\sigma_z}_1|\hat{\dot{x}}|\psi^{\sigma_z}_{-2}\rangle=0,\\
\langle\psi^{\sigma_z}_2|\hat{\dot{y}}|\psi^{\sigma_z}_{-2}\rangle=&\langle\psi^{\sigma_z}_1|\hat{\dot{y}}|\psi^{\sigma_z}_{-2}\rangle=0,\\
\langle\psi^{\sigma_z}_2|\hat{\dot{z}}|\psi^{\sigma_z}_{-2}\rangle=&i\frac{z_0}{\hbar}\Delta E,\;\;\;\langle\psi^{\sigma_z}_1|\hat{\dot{z}}|\psi^{\sigma_z}_{-2}\rangle=i\sigma_z\frac{z_0}{\hbar}\alpha k\sqrt{1-\eta_{\boldsymbol{k}}}.\label{eq_excitation_2}
\end{split}
\end{align}
Here, although the matrix components depend on the spin index $\sigma_z$ explicitly, the spin dependence does not appear in the optical conductivity.
This is because the amplitude of the component contribute to Eq.\;(\ref{eq_dynamical_conductivity}).
These zero components obviously show the absence of optical conductivity associated to specific pairs of conduction and valence bands under some conditions of photon polarization.

%%%%%%%%%%%%%%%%%%%%%%%%%%%%%%%%%%%%%%%%%%%%%%

The analytic representation of the dynamical polarization matrix indicates the presence of restrictions to electronic excitation under the symmetrical condition of $M_2$HfC$_2$O$_2$.
Since the zero components indicate the absence of dynamical polarization along the corresponding axis between the two bands, the incident photons linearly polarized along the axis induce no excitation between these bands.
For instance, the excitation from the lowest valence band is prohibited for the parallel polarization direction to the layer, i.e., the $xy$-plane.
On the other hand, photons polarized along the perpendicular direction to the layer cannot induce the excitation from the highest valence band.
These restrictions to the excitation process describe the absent peaks of the optical conductivity spectrum especially for $\theta=0^\circ$ and $90^\circ$ in Fig.\;\ref{fig_dynamical_conductivity}.
Actually, two characteristic variations \textcircled{1} and \textcircled{2} are absent in the spectrum for the polarization angle $\theta=90^\circ$ since these peaks are associated with the excitation from the highest valence band.
In contrast, the other two features \textcircled{3} and \textcircled{4} cannot be observed for $\theta=0^\circ$. 
Therefore, the analytic representation of the dynamical polarization matrix describes the fundamental nature of the dynamical response and enables us to investigate the dynamical electronic properties from a microscopic point of view.

The effective model also reveals the irrelevance of the electronic states at the $\Gamma$ point in the case of the lowest-energy peak of the optical conductivity in Fig.\;\ref{fig_dynamical_conductivity}.
In the case of the lowest-peak, the relevant electronic excitation is obviously that between the highest valence band and the lowest conduction band, and its contribution can be evaluated by the dynamical polarization matrix in Eq.\;(\ref{eq_excitation_1}).
The dynamical polarization matrix components $\langle \psi_1^{\sigma_z}|\hat{\dot{x}}|\psi_{-1}^{\sigma_z}\rangle$ and $\langle \psi_1^{\sigma_z}|\hat{\dot{y}}|\psi_{-1}^{\sigma_z}\rangle$ are proportional to the wave number $k$ and thus absent at the $\Gamma$ point.
Since peak structures in the spectra of optical conductivity are associated with the local minima or maxia in the energy dispersion, the irrelevance of the excitation at the $\Gamma$ point indicates that the local minima of the lowest conduction band around the $\Gamma$ point are relevant to the lowest-energy peak structure.
Actually, the JDOS peak attributed to the local mimima appears below the excitation energy at the $\Gamma$ point as shown in Fig.\;\ref{fig_JDOS}.
The irrelevance of the $\Gamma$ point can also be observed as the slight deviation between the excitation energy at the $\Gamma$ point and the peak position in Fig.\;\ref{fig_dynamical_conductivity}.
Therefore, the effective model gives a clear description oft the absence of double-peak structure in the optical conductivity spectrum attributed to the local maximum and minima of the lowest conduction band.

%%%%%%%%%%%%%%%%%%%%%%%%%%%%%%%%%%%%%%%%%%%%%%

\subsection{Parameters for Realistic Materials}

%%%%%%%%%%%%%%%%%%%%%%%%%%%%%%%%%%%%%%%%%%%%%%

\begin{figure}[htbp]
\begin{center}
 \includegraphics[width=80mm]{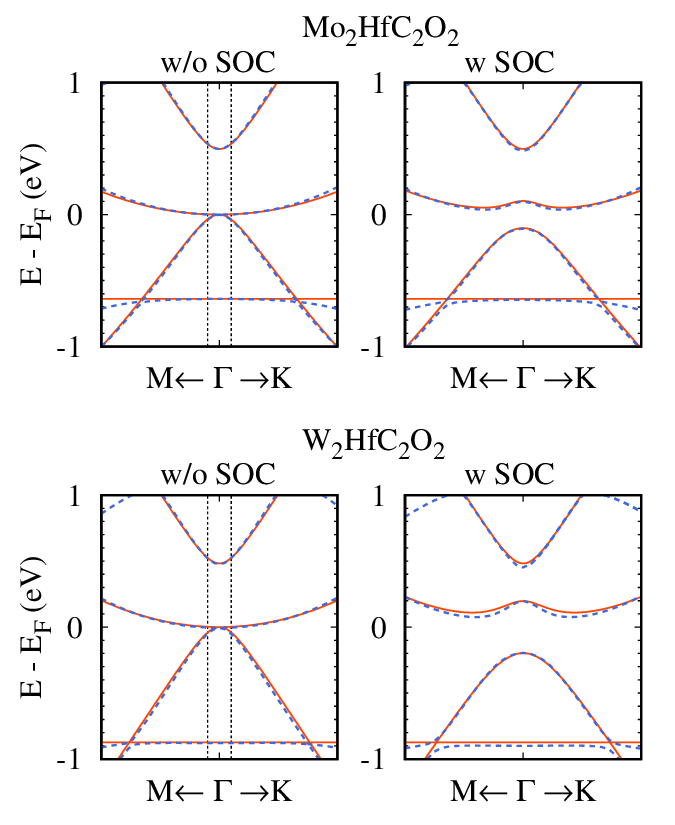}
\caption{
The band structure reproduced by the effective model for Mo$_2$HfC$_2$O$_2$ and W$_2$HfC$_2$O$_2$. The reproduced dispersion and the first-principles are given by the solid curves and the dashed curves, respectively. The left and right panels present the dispersion without SOC and that with SOC, respectively. The vertical dashed lines indicate the region referred for the calculation of the effective masses in table \ref{tab_effective_mass}.
 }\label{fig_effective_band}
\end{center}
\end{figure}

Finally, the parameters in the effective model are estimated to describe the specific materials in terms of the band structure around the $\Gamma$ point.
The effective model contains two energy constants, $E_t$ and $E_b$, and four coefficients, $\alpha$, $\beta_1$, $\beta_2$, and $\gamma$, associated with the energy dispersion.
The two energy constants can be obtained referring to the band edges at the $\Gamma$ point without SOC.
The other parameters can be determined from the effective mass of the four bands.
In the vicinity of the $\Gamma$ point, i.e., $\cos\phi_{\boldsymbol{k}}\simeq1$, the energy dispersion can be given by the summation of $E_j^{(0)}+\beta_{|j|}k^2$, where the subscript of $\beta_{|j|}$ is attributed to the equivalence of the parameters between the conduction and valence bands as shown in Eq.\;(\ref{eq_k-dependence}).
Under the same condition, $\Delta_k$ can also be approximated by a quadratic form, $\Delta_k\simeq(\alpha/E_t)^2k^2$.
Then the effective mass for each band is represented by
\begin{align}
\begin{split}
\frac{1}{m_2}=&\frac{2}{\hbar^2}\left(\frac{\alpha^2}{E_t^2}+\beta_2\right),\\
\frac{1}{m_1}=&\frac{2}{\hbar^2}\left(\gamma+\beta_1\right),\\
\frac{1}{m_{-1}}=&\frac{2}{\hbar^2}\left(-\gamma-\frac{\alpha^2}{E_t^2}+\beta_1\right),\\
\frac{1}{m_{-2}}=&\frac{2}{\hbar^2}\beta_2.
\end{split}
\end{align}
The effective masses are estimated from the first-principles band without SOC in the range of wave number $k<0.025\times\pi/a$ around the $\Gamma$ point.
In this range, the dispersion of the lowest band is quite small in comparison with the other three bands, and thus it is assumed to be flat, i.e., $\beta_2=0$.
The other parameters to reproduce the first-principles band are presented in Table \ref{tab_effective_mass}. 
The reproduced band structures are also presented in Fig.\;\ref{fig_effective_band} with the first-principles bands for Mo$_2$HfC$_2$O$_2$ and W$_2$HfC$_2$O$_2$ in both cases with SOC and without SOC.
The similarity of the band structures in the effective model and first-principles calculations shows the validity of the effective model for theoretical analyses of the low energy electronic properties of Mo$_2$HfC$_2$O$_2$ and W$_2$HfC$_2$O$_2$.

\begin{table}
\caption{The parameters for the effective model to simulate actual materials, Mo$_2$HfC$_2$O$_2$ and W$2$HfC$_2$O$_2$. The two energy constants $E_t$ and $E_b$ are given in the unit of eV. The units for other parameters are eV$\cdot$\AA$^2$ for $\alpha$, and eV$\cdot$\AA\; for $\beta_j$ and $\gamma$. The SOC energy $\lambda$ is also presented in eV.
}
\begin{ruledtabular}
\begin{tabular}{c c c c c c c}
&$E_t$&$E_b$&$\alpha$&$\beta_1$&$\gamma$&$\lambda$\\ \hline
Mo$_2$HfC$_2$O$_2$&0.497&-0.639&3.892&2.898&-0.379&0.103 \\ 
W$_2$HfC$_2$O$_2$&0.482&-0.873&3.974&1.712&1.223&0.197\\ 
\end{tabular}\label{tab_effective_mass}
\end{ruledtabular}
\end{table}

The parameters of the effective model are also consistent with the numerical results of optical conductivity for $\theta=0^\circ$ in terms of the amplitude in the two compounds, Mo$_2$HfC$_2$O$_2$ and W$_2$HfC$_2$O$_2$.
For $\theta=0^\circ$, the spectrum possesses two characteristic structures, a peak \textcircled{1} and a step-like increment \textcircled{2}.
The peak structure is attributed to the excitation between the highest valence band and the lowest conduction band, and thus the amplitude depends on the dynamical polarization matrix component between $|\psi_{1}^{\sigma_z}\rangle$ and $|\psi_{-1}^{\sigma_z}\rangle$ in the $xy$-plane.
Since the amplitude of the matrix component is proportional to a coefficient $\gamma$ in Table \ref{tab_effective_mass}, W$_2$HfC$_2$O$_2$ gives a much larger value than that in Mo$_2$HfC$_2$O$_2$.
Then the difference in the coefficient can describe the larger difference of peak height between W$_2$HfC$_2$O$_2$ and Mo$_2$HfC$_2$O$_2$.
The step-like increment exhibits no significant difference in the height for the two compounds and it can also be described by the similarity of a coefficient $\alpha$ in the two compounds because the dynamical polarization matrix component $\langle\psi_{2}^{\sigma_z}|\hat{\dot{x}}|\psi_{-1}^{\sigma_z}\rangle$ or $\langle\psi_{2}^{\sigma_z}|\hat{\dot{y}}|\psi_{-1}^{\sigma_z}\rangle$ is responsible for the relevant excitation process.

%%%%%%%%%%%%%%%%%%%%%%%%%%%%%%%%%%%%%%%%%%%%%%

\section{discussion and conclusion}\label{sec_conclusion}

%%%%%%%%%%%%%%%%%%%%%%%%%%%%%%%%%%%%%%%%%%%%%%

In general, optical properties are dominated by electronic states extended in the bulk of material and it is not affected drastically by the edge states, the characteristic feature of topological insulators.
However, the topologically-nontrivial properties are attributed to the electronic structure in the bulk, e.g., the parity inversion between conduction and valence bands at high symmetry points in the Brillouin zone.\cite{Fu2007}
Actually, the odd-parity and even-parity orbitals are switched gradually with the wave number on the highest valence band and the second lowest conduction band as shown in Fig.\;\ref{fig_orbital_amplitude}.
In the effective model, the parity inversion can be described by the off-diagonal components proportional to the constant $\alpha$ in Eq.\;(\ref{eq_k-dependence}).
The mixing term also introduces the $\boldsymbol{k}$-dependent relative phase difference, so-called chiral phase, among the Wannier orbitals in the electronic states.
In the case of the optical properties, the parity inversion leads to a significant effect on the spectrum of optical conductivity especially for the step-like increments, e.g., \textcircled{2}, in Fig.\;\ref{fig_dynamical_conductivity}.
Actually, the analytic representation of the dynamical polarization, $\langle\psi_{2}^{\sigma_z}|\hat{\dot{x}}|\psi_{-1}^{\sigma_z}\rangle$ or $\langle\psi_{2}^{\sigma_z}|\hat{\dot{y}}|\psi_{-1}^{\sigma_z}\rangle$, reveals that it is proportional to the coupling constant $\alpha$ between the opposite parity orbitals, and it remains non-zero even at the $\Gamma$ point.
The constant dynamical polarization element results in the steep variation of optical conductivity.
Therefore, the step-like increments can be evidence of the parity inversion associated with the topologically-nontrivial electronic structure in these MXenes.

In conclusion, it is theoretically revealed that the optical conductivity of Mo$_2$HfC$_2$O$_2$ and W$_2$HfC$_2$O$_2$ exhibits several characteristic features, peaks and step-like increments, associated with topologically unconventional electronic states in the low photon energy region.
The relevant bands to the dynamical response in this region are two conduction bands and two valence bands which are also responsible for producing the topological nature, i.e., parity inversion.
The numerical calculations present a fascinating photon polarization angle dependence, switching of the presence and the absence of the features in the optical conductivity spectrum, between the perpendicular axis and the parallel axis to the layer.
The effective model analysis is also performed to investigate the origin of the characteristic features in the spectrum.
A minimal model is generated referring to the crystal symmetries and applied to derive the analytic formula of the dynamical electronic polarization matrix associated with the optical conductivity.
The analytic representation reveals that the polarization angle dependence is strongly related to parity inversion between the conduction and valence bands, the key feature of the topological phase.
The fitting parameters for the first-principles band structure are also provided and they also describe the relative amplitudes of optical conductivity between Mo$_2$HfC$_2$O$_2$ and W$_2$HfC$_2$O$_2$.
The reproducibility of the band structure and the consistency about the amplitude suggest the validity of the effective model for the investigation of the electronic properties in topologically nontrivial MXenes.

\begin{acknowledgements}
This work was supported by JSPS KAKENHI Grant Number JP23K03289.
\end{acknowledgements} 

\bibliography{Topological_MXene}

\end{document}